\documentclass{SciPost}

\hypersetup{
    colorlinks,
    linkcolor={red!50!black},
    citecolor={blue!50!black},
    urlcolor={blue!80!black}
}

\usepackage[utf8]{inputenc} 
\usepackage{amsmath, amssymb, amsthm} 
\usepackage{amsfonts}
\usepackage{appendix}
\usepackage{cite} 
\usepackage{dsfont}
\usepackage{graphicx}
\usepackage{adjustbox}
\usepackage{float} 
\usepackage{hyperref}
\usepackage{geometry} 
\usepackage{enumitem} 
\usepackage{fancyhdr}
\usepackage{subcaption}
\usepackage{physics}
\usepackage[T1]{fontenc}
\usepackage[utf8]{inputenc}
\usepackage[polish,english]{babel}
\usepackage{xcolor}
\usepackage[bitstream-charter]{mathdesign}
\usepackage[normalem]{ulem}

\DeclareSymbolFont{usualmathcal}{OMS}{cmsy}{m}{n}
\DeclareSymbolFontAlphabet{\mathcal}{usualmathcal}

\fancypagestyle{SPstyle}{
\fancyhf{}
\lhead{\colorbox{scipostblue}{\bf \color{white} ~SciPost Physics }}
\rhead{{\bf \color{scipostdeepblue} ~Submission }}

\fancyfoot[C]{\textbf{\thepage}}
}

\newcommand{\ham}{\mathcal{H}}
\newcommand{\agp}{\mathcal{A}_\lambda}
\newcommand{\agpe}{\mathcal{A}_\epsilon}
\newcommand{\agpb}{\mathcal{A}_\beta}
\newcommand{\gradagp}{\nabla \agp}
\newcommand{\gradagpe}{\nabla \agpe}
\newcommand{\gradagpb}{\nabla \agpb}
\renewcommand{\pb}[2]{\left\{#1, #2\right\}}
\newcommand{\pert}{V}
\newcommand{\G}{\hat{\mathcal{G}}_{\lambda}}
\newcommand{\bigO}{\mathcal{O}}
\newcommand{\pertt}{\pert(t)}
\newcommand{\dpdq}{\begin{pmatrix} \delta p \\ \delta q \end{pmatrix}}
\newcommand{\dpdqt}{\begin{pmatrix} \delta p(t) \\ \delta q(t) \end{pmatrix}}
\newcommand{\dpdqz}{\begin{pmatrix} \delta p(0) \\ \delta q(0) \end{pmatrix}}
\newcommand{\Id}{\mathds{1}}
\newcommand{\Symp}{\mathcal{I}}
\newcommand{\minus}{\text{-}}
\newcommand{\blockmatrix}[4]{
  \big(\begin{smallmatrix}
    #1 & #2 \\
    #3 & #4
  \end{smallmatrix}\big)
}

\begin{document}

\pagestyle{SPstyle}

\begin{center}{\Large \textbf{\color{scipostdeepblue}{
Gradient of the Adiabatic Gauge Potential in Classical Systems\\
}}}\end{center}

\begin{center}\textbf{
Nathan Rose\textsuperscript{1$\star$}, 
Nachiket Karve\textsuperscript{1} and
David K. Campbell\textsuperscript{1}
}\end{center}

\begin{center}
{\bf 1} Department of Physics, Boston University, Boston, Massachusetts 02215, USA
\\[\baselineskip]
$\star$ \href{mailto:email1}{\small nrose68@bu.edu}
\end{center}

\section*{\color{scipostdeepblue}{Abstract}}
\textbf{\boldmath{%
The adiabatic gauge potential (AGP) is the generator of unitary transformations which preserve the eigenbasis of a quantum Hamiltonian under parametric variation. While its usefulness in quantum mechanics has been thoroughly demonstrated in recent years, less attention has been given to its behavior in classical systems, where the AGP is a phase space function and its gradient defines special canonical transformations. In this paper we propose an efficient method to compute the gradient of the AGP as a classical function. We demonstrate that the obtained canonical transformation reproduces expected results for simple orbits and integrable systems for which the adiabatic limit is well-defined. In chaotic systems the gradient diverges in a way that is related to Lyapunov times. 
}}

\vspace{\baselineskip}



\vspace{10pt}
\noindent\rule{\textwidth}{1pt}
\tableofcontents
\noindent\rule{\textwidth}{1pt}
\vspace{10pt}

\section{Introduction}
\label{sec:intro}
Many physical systems are modeled by a Hamiltonian with time-varying parameters. When the rate of variation is infinitesimally small, the system undergoes adiabatic evolution. In quantum mechanics, the adiabatic theorem \cite{Ka1950} states that a system prepared initially in the eigenstate of a gapped Hamiltonian will remain in an instantaneous eigenstate under adiabatic evolution. 
Eigenstates of the Hamiltonian at different values of the parameter are related by a unitary transformation, and this unitary transformation is generated by a Hermitian operator known as the adiabatic gauge potential (AGP) \cite{KoSeMePo17}. The AGP has received significant attention in recent years in the contexts of state preparation and shortcuts to adiabaticity \cite{DeRi03, Be09, Ja2013, dCa13, HoRaJa25}, in probing chaos and integrability breaking in quantum and classical systems \cite{PaClCaPoSe20, SePo21, SePo23, LiMaKiPoFl24, SwLyVi25, KaRoCa25}, and Born-Oppenheimer forces \cite{BaArChPo25}.

Exact solutions for the AGP are in general difficult to come by. In generic systems, the AGP is a nonlocal operator which additionally hinders its feasibility in experiments. This has motivated the development of local approximations to the AGP based on variational schemes \cite{KoSeMePo17, SePo17, GjCaPo22}, Krylov expansions \cite{TadCa24, Bh23}, and polynomial representations \cite{FiNoCaLuSe25, MoPo25}.

Recent works have established that both quantum and classical chaos can be quantified with respect to a system's sensitivity to adiabatic deformations \cite{LiMaKiPoFl24, KaRoCa25, KaRoCa2025_2}. As a probe of chaos, the object generally studied is the regularized norm of the AGP, which is defined via a regularization parameter which acts as a long-time cutoff. The scaling of the AGP norm when the cutoff time is large distinguishes integrable and chaotic dynamics. Perhaps surprisingly, the AGP norm grows fastest in systems where integrability is only weakly broken. Open questions remain, however, about how this measure which probes the low frequency behavior of the spectral function is related to other conventional measures of chaos based on short time dynamical instabilities, such as out-of-time-order correlators (OTOCs) \cite{KuGrPr17, GuKiZh22, Tr23} and operator growth rates \cite{PaCaAvScAl19}.

Much of the focus in this area has been with quantum systems, though it has been pointed out that the AGP formalism carries over to classical mechanics by taking the semiclassical limit -- replacing commutators with Poisson brackets and operators with functions in phase space. Pioneering works by Robbins and Berry \cite{RoBe1992}, and Jarzynski \cite{Ja1995} pointed out the difficulty in defining the AGP or related objects in chaotic classical systems. Nevertheless, having additional tools to study the AGP in classical systems could lead to a better understanding of how to design local approximations or understand its role as a probe of chaos.

In classical mechanics, any differentiable function generates infinitesimal canonical transformations via Hamilton's equations. These involve first order derivatives of the considered function in all coordinate directions, and therefore the gradient of a function determines the canonical transformations it generates. Thus, the gradient of the AGP, which is not defined in quantum mechanics, describes how to adiabatically deform classical trajectories. Sensitivity of individual trajectories to perturbation is clearly a hallmark sign of chaos in classical systems which likewise has no direct quantum analog.

In this paper, we propose an efficient method of directly computing the gradient of the AGP. This method involves the evolution of a trajectory and the linearized flow of tangent vectors around it, and is amenable to numerical evaluation. In Sec. \ref{sec:theory} the AGP is reviewed in quantum and classical mechanics. In Sec. \ref{sec:methods} the method for computing the gradient of the AGP is described. In Sec. \ref{sec:examples} we show that our calculations reproduce correct results in simple scenarios. In Sec. \ref{sec:results}  numerical results are presented for both systems with integrable and chaotic dynamics. Finally, concluding remarks are given in Sec. \ref{sec:discussion}.

\section{Adiabatic Gauge Potential}
\label{sec:theory}
In this section, we briefly review how the adiabatic gauge potential is defined in quantum mechanics and discuss the classical picture obtained by taking the semiclassical limit. Further details can be found in \cite{KoSeMePo17}.
\subsection{Quantum Mechanics}

In quantum mechanics, the AGP is defined by considering the change to an eigenstate $\ket{n(\lambda)}$ as a parameter $\lambda$ of the Hamiltonian $\hat\ham(\lambda)$ is varied \cite{KoSeMePo17, LiMaKiPoFl24}, such that
\begin{equation}
    \braket{m(\lambda)}{n(\lambda + \delta\lambda)} \approx \delta_{mn} - \frac{i}{\hbar}\ \bra{m(\lambda)} \hat{\agp}\ket{n(\lambda)}\ \delta \lambda + \mathcal{O}(\delta\lambda^2)
\end{equation}

The off-diagonal elements of $\hat\agp$ are given by first-order perturbation theory 

\begin{equation}
    (\hat{\agp})_{mn} \equiv -i \frac{\bra{m}\hat{V}\ket{n}}{\omega_{mn}}
    \label{eq:agp_matrix_element}
\end{equation}
where $\hat{V}\equiv \partial_\lambda \hat{H}$ and $\omega_{mn}$ denotes the difference in energy levels $\hbar\omega_{mn} = E_m - E_n$. The diagonal elements of $\hat\agp$ are arbitrary; they correspond to the ability to define eigenstates up to an overall phase. The AGP can be written in a way that makes this gauge freedom manifest 

\begin{equation}
    \left[\hat\agp, \hat\ham\right] = \hat{V} - \G
    \label{eq:quantum_agp_commutator}
\end{equation}
where $\G$ is defined $\G\equiv \sum_n \hat{V}_{nn}\ \ket{n}\bra{n}$ and $\hat{V}_{nn}\equiv \bra{n}\hat{V}\ket{n}$. It is clear that the $\agp$ is defined up to an operator commuting with the Hamiltonian. 

To keep Eq. \eqref{eq:agp_matrix_element} well defined in the case of accidental (near) degeneracies, one can introduce a parameter $\mu$ such that the regularized matrix elements are 
\begin{equation}
    \hat{\agp}(\mu)_{mn} = -i \left(\frac{\omega_{mn}}{\omega_{mn}^2 + \mu^2}\right) \bra{m}\hat{V}\ket{n} 
    \label{eq:agp_reg_matrix_element}
\end{equation}

The physical meaning of $\mu$ will be described in the next section. A formal solution for an operator satisfying Eq. \eqref{eq:agp_reg_matrix_element} is

\begin{equation}
    \hat{\agp}(\mu) = -\frac{1}{2} \int dt\ e^{\minus\mu|t|}\ sgn(t)\ \hat{V}(t)
\end{equation}

where the operator $\hat{V}(t)$ evolves under Heisenberg evolution. Similarly, the operator $\G$ can be defined with respect to $\mu$ as 
\begin{equation}
    \G(\mu) = \frac{\mu}{2}\int_{\minus \infty}^{\infty} dt\ e^{\minus\mu|t|}\ \hat{V}(t)
\end{equation}
which is now no longer exactly diagonal but has suppressed matrix elements between eigenstates within energy difference greater than $\hbar \mu$.

\subsection{Classical Mechanics}
Following \cite{Ja1995}, the semiclassical limit of Eq. \eqref{eq:quantum_agp_commutator} is taken to obtain a classical expression for the AGP as a phase space function $\agp(p,q)$
\begin{equation}
    \pb{\agp}{\ham} = V - \langle V\rangle
    \label{eq:cl_pb_agp_def}
\end{equation}

where the brackets denote infinite time average taken over a trajectory passing through phase space point $(p,q)$. Under this condition \eqref{eq:cl_pb_agp_def} the AGP is defined only up to any conserved function $f(p,q)$.  This ambiguity is removed by fixing the average value 

\begin{equation}
    \langle \agp \rangle=0.    
    \label{eq:agp_avg_zero}
\end{equation}

Reversing the order in the Poisson bracket of Eq. \ref{eq:cl_pb_agp_def} gives $\dot{\agp} = \pb{\ham}{\agp}$, which may be used to obtain a finite time representation of $\agp$
\begin{equation}
    \agp(t) = \agp(0)\ - \int_0^t dt\ V(t) - \langle V\rangle
    \label{eq:time_rep}
\end{equation}

where it is important to note that $V(t) \equiv V(p(t), q(t))$ is evaluated along a dynamical trajectory of the system.

As a function of time, the value of $\agp(t)$ is analogous to the diffusion of a particle in one dimension, with velocity $V - \langle V\rangle$. Different classes of diffusion correspond to different kinds of dynamics, with bounded motion occurring for integrable systems and normal diffusion occurring for chaotic systems. Interestingly, in systems where integrability is weakly broken, $\agp(t)$ exhibits (superlinear) anomalous diffusion, at least transiently over a long thermalization time \cite{KaRoCa25}. 

The fact that in general the time representation $\agp(t)$ diffuses unboundedly while the trajectory remains located in a finite region of phase space demonstrates the difficulty in defining the AGP for all times. To treat $\agp(p,q)$ as a phase space function, one can formally solve Eqs. \eqref{eq:cl_pb_agp_def}, \eqref{eq:agp_avg_zero} with 

\begin{equation}
    \agp(p, q) = \lim_{\mu\rightarrow 0}\ \agp(p, q; \mu)
    \label{eq:agp_limit}
\end{equation}
where 
\begin{equation}
    \agp(p, q; \mu) = -\frac{1}{2}\int_{\minus \infty}^{\infty} dt\ e^{-\mu t}\ sgn(t)\ \pertt
    \label{eq:agp_general}
\end{equation} 

Again the AGP diverges as $\mu\rightarrow 0$ in chaotic systems and so is ill-defined. Therefore, it is better to consider as the primary object of interest not Eq. \eqref{eq:agp_limit} but the `regularized' AGP $\agp(p, q;\mu)$, where $\mu$ is a physical parameter representing a cutoff time suppressing long-time correlations of the perturbation $V$. As a function of $\mu$, the AGP is now interpreted as generating canonical transformations that preserve trajectories under deformation of the Hamiltonian up to a time scale $\mu^{\minus 1}$.

Taking the Poisson bracket of $\agp$ with a given function $f(p, q)$ yields

\begin{equation}
    \pb{\agp}{f} = - \gradagp\ \Symp\ \nabla f
    \label{eq_Af_poisson_bracket}
\end{equation}

where the $2N\times2N$ symplectic matrix $\Symp$ written as a block matrix in terms of the $N\times N$ identity matrix is $\Symp = \blockmatrix{0}{\minus \Id}{\Id}{0}$. Characteristics of this equation are generated with $\agp$ playing the role of (parameter dependent) Hamiltonian 

\begin{equation}
    \begin{pmatrix} p \\ q \end{pmatrix}_{\lambda + d\lambda} = \begin{pmatrix} p \\ q \end{pmatrix}_\lambda + d\lambda\ \ \Symp\ \gradagp
    \label{eq:pq_trans}
\end{equation}

such that

\begin{equation}
    f_{\lambda + d\lambda} \equiv f(p_{\lambda + d\lambda}, q_{\lambda + d\lambda}) = f_{\lambda} + d\lambda \pb{\agp}{f_\lambda}
\end{equation}

In this way $\agp$ and its gradient $\gradagp$ generate a canonical transformation.  If $f$ is nonnegative and bounded then we can interpret it as a density of phase space points each evolving according to the Hamiltonian.  Furthermore, if the density $f$ is stationary under an original Hamiltonian $\mathcal{H}_\lambda$, then $f_{\lambda + d\lambda}$ is stationary under the perturbed one $\mathcal{H}_{\lambda + d\lambda}$. In this sense we can consider Eq. \eqref{eq:pq_trans} as continuously deforming trajectories among a family of Hamiltonians.

\section{Methods}
\label{sec:methods}

\subsection{Algorithm for Computing the AGP Gradient}

The classical expression Eq. \eqref{eq:agp_general} can be evaluated by selecting an initial condition $(p, q)$ and inserting the solution $(p(t), q(t))$ into the integral. In general, when a closed form solution cannot be obtained, this expression may be evaluated by numerically integrating the equations of motion from the initial point $(p,q)$.

With proper regularization, the AGP can be expanded locally as

\begin{equation}
    \agp(p_0 + \delta p,\ q_0 + \delta q) \approx \agp(p_0,q_0) +  \gradagp \cdot \dpdq
\end{equation}

Consider Eq. \eqref{eq:agp_general} for a perturbed initial condition $(p + \delta p, q + \delta q)$. Expanding the function $V(t)$ inside the integral, we have 

\begin{equation}
\begin{aligned}
    V(p(t) +\delta p(t),\ q(t) + \delta q(t)) &\approx V(p(t), q(t)) + \nabla V(p(t), q(t)) \cdot \dpdqt + \bigO(\delta^2) \\
    &\approx V(t) + \delta\ \nabla V(t) \cdot \xi(t) + \bigO(\delta^2)
\end{aligned}
\end{equation}

where $\xi(t) = \dpdqt / \left\lVert \dpdqz \right\rVert$ and $\delta \equiv \left\lVert \dpdqz \right\rVert$. In the limit $\delta\rightarrow0$, $\xi(t)$ is described by a tangent vector at point $(p(t), q(t))$ and is evolved according to the linearized equations of motion. To find the gradient $\gradagp$, initialize an orthonormal basis of tangent vectors $\Phi$, satisfying

\begin{equation}
\begin{aligned}
    &\dot\Phi(t) =  J\ \Phi(t) \\
    &\Phi(0) = \Id
\end{aligned}
\end{equation}
With $J_{ij} = \Symp\ \frac{\partial^2 \ham}{\partial x_i \partial x_j} $, and $x_i$ is used to denote both $p_n$ and $q_n$. Together, we have the formula for $\gradagp$

\begin{equation}
    \gradagp(p,q; \mu) = -\frac{1}{2} \int_{\minus \infty}^{\infty}dt\ \text{sgn}(t)\ e^{\minus \mu |t|}\ \nabla V(t)\cdot\Phi(t) 
    \label{eq:agp_grad}
\end{equation}
This expression can be evaluated analytically only in special cases but is amenable to numerical treatment using standard methods \cite{CaSa1993}. Showing how $\gradagp$ can be computed using Eq. \eqref{eq:agp_grad} and interpreting the results comprise the bulk of this paper.

\subsection{Periodic Orbit}

In the case of a periodic orbit, the gradient of the AGP can be neatly represented in terms of the monodromy matrix $F$ \cite{YaSt75}, whose eigenvalues and associated eigenvectors determine which directions of $\gradagp$ may be convergent or divergent.

Consider a periodic orbit $(p(t),\ q(t))$ with period $T$ such that $(p(t+T),\ q(t+T)) = (p(t),\ q(t))$. In this case, the time integral in Eq. \eqref{eq:agp_grad} can be represented as a sum of single-period integrals

\begin{equation}
    \int_0^{\infty} dt\ e^{\minus\mu t} \nabla V(t)\ \cdot\ \Phi(t) = \sum_{j=0}^{\infty}  \int_{jT}^{(j+1)T} dt\ e^{\minus\mu t}\ \nabla V(t)\ \cdot\ \Phi(t)
\end{equation}
Using the periodicity of the orbit, we note that $\nabla V(t+jT) = \nabla V(t)$, whereas, the tangent vectors after $j$ number of periods are given by
\begin{equation}
    \Phi(t + jT) = \Phi(t)\ F^j.
\end{equation}
Together we get
\begin{equation}
     \int_{0}^{T} dt\ e^{\minus\mu t}\ \nabla V(t)\ \cdot\ \Phi(t)\cdot \sum_{j=0}^{\infty}\  (e^{\minus\mu T} F)^j
\end{equation}
If the orbit is stable, then the eigenvalues of $F$ will all have magnitude 1, i.e. they will lie on the unit circle in the complex plane. In this case, the sum converges as a geometric series for arbitrary finite $\mu$, and the result is

\begin{equation}
    \Bigg[\int_{0}^{T} dt\ e^{\minus\mu t}\ \nabla V(t)\ \cdot\ \Phi(t) \Bigg] \big(\Id - e^{\minus\mu T} F\big) ^{\minus 1}
    \label{eq:periodic_forward_term}
\end{equation}
The integral over negative times in Eq. \eqref{eq:agp_grad} gives a second similar but not identical term

\begin{equation}
     \Bigg[\int_{\minus T}^{0} dt\ e^{\mu t}\ \nabla V(t)\ \cdot\ \Phi(t) \Bigg] \big(\Id - e^{\minus\mu T}F^{\minus 1}\big) ^{\minus 1}
\end{equation}

Thus $\gradagp$ on a periodic orbit can be written in terms of the monodromy matrix, and time integration need only be performed over a single period. Interestingly, a matrix factor like $\big(\Id - F\big) ^{\minus 1}$ appears in other contexts as well, such as Newton-Raphson minimization methods used for finding periodic orbits \cite{FlIvKa06}.

\subsection{Chaotic Dynamics}

For a generic orbit, a generalization of the eigenvectors of the Monodromy matrix $F$ is given by covariant Lyapunov vectors (CLVs) \cite{GiPoTuChPo2007, KuPa2012}. In the case of a chaotic system with a nondegenerate Lyapunov spectrum $\lambda_1>\lambda_2>...>\lambda_N=0$, CLVs are position dependent unit vectors $v_j$ which define invariant one dimensional subspaces of the tangent space. A vector parallel to a covariant vector $v_j$ at time $t=0$ will remain in that subspace and grow asymptotically as $e^{\lambda_jt}$ on average. See App. \ref{app:clvs} for further details.

The CLVs determine directions in which $\gradagp(\mu)$ diverges for finite $\mu < \lambda_j$. We may still consider the action of $\agp(\mu)$ on a smooth phase space distribution $\rho(p, q)$ . If we demand that $\rho$ be transformed continuously by $\agp$, then this places some restrictions on $\rho$. In particular, if we demand that 

\begin{equation}
    \pb{\agp(\mu)}{\rho} = \frac{\partial\agp}{\partial p}\frac{\partial\rho}{\partial q} - \frac{\partial\agp}{\partial q}\frac{\partial\rho}{\partial p} = -\big(\gradagp \Symp\ \nabla \rho)
    \label{eq:rho_pb}
\end{equation}

is finite, then for each direction $v_j$ of $\gradagp$ that diverges as $\mu \rightarrow 0$, there corresponds an orthogonal direction $u_j$ such that $\nabla\rho \cdot u_j=0$. For a generic system without additional conservation laws there will be $2N-2$ directions in which the gradient of $\rho$ must vanish for Eq. \eqref{eq:rho_pb} to be finite for arbitrarily small $\mu$. Indeed, this is satisfied for microcanonical or Boltzmann distributions which are flat over an energy shell.

\subsection{Alternate Regularization Schemes}

As mentioned previously, typical infinitesimal perturbations in chaotic systems grow exponentially, so that $|\Phi(t)|\sim e^{\Lambda t}$, where $\Lambda$ is the maximal Lyapunov exponent (MLE), and from Eq. \eqref{eq:agp_grad} we find that derivatives of $\agp$ in generic directions diverge for $\mu < \Lambda$. For the AGP norm, a regularization factor of $e^{\minus\mu |t|}$ guarantees convergence for any finite $\mu$, but this is not true for the gradient. 

Note that the factor of $e^{\minus\mu |t|}$ appearing in Eqs. \eqref{eq:agp_general} and $\eqref{eq:agp_grad}$ is a particular choice of a filter function supressing the long time behavior of the integrand \cite{ha10}. In general, we could consider

\begin{equation}
    \agp(p, q; \mu) = -\frac{1}{2} \left[ \int_{\minus \infty}^{\infty} dt\ g(|\mu t|)\ sgn(t)\  \pertt \right] 
    \label{eq:agp_general_filter}
\end{equation} 
with $g(x)$ sufficiently small when $|x|$ is large. To ensure the expression for $\gradagp(\mu)$ converges for any $\mu$, we need a filter function $g$ that decays faster than exponentially. A hard cutoff $g(x) =\Theta(1 - x)$ where $\Theta$ is the Heavyside step function, is a valid option but the resulting value of $\gradagp(\mu)$ can depend strongly on $\mu$. An alternative is to use a Gaussian function $g(x) = e^{\minus x^2}$, which decays sufficiently rapidly to guarantee convergence. This is the regularization scheme used for numerical results in chaotic systems in this paper. Ultimately, the choice is not very important, since whether $\gradagp$ is infinite or merely exponentially large makes little difference in practice. In cases such as integrable systems where $\gradagp(\mu)$ converges as $\mu\rightarrow0$, all three choices of filter function yield consistent results. See App. \ref{app:regularization} for further details.

\section{Analytical Examples}
\label{sec:examples}
In this section we consider Eq. \ref{eq:agp_grad} analytically in some simple scenarios.

\subsection{Fixed Point}
The simplest state to consider is one that remains constant for all times, i.e. a fixed point $(p_0, q_0)$ of some system such that $(p(t), q(t)) = (p_0, q_0)$ for all $t$. Since the perturbing function $V(t)=V(0)$ is constant in time, we have that $\agp=0$ according to Eq. \eqref{eq:agp_general}.

We consider a system with one degree of freedom for simplicity, though generalization to higher dimensions is straightforward. Supposing that the fixed point is stable, we can consider as the base Hamiltonian a harmonic oscillator, and add a perturbation $V(q)$
\begin{equation}
    \ham = \frac{1}{2}\big(p^2 + \omega^2 q^2\big) + \epsilon\ V(q)
    \label{eq:perturbed_ho}
\end{equation}
so that the fixed point lies at $(p_0, q_0) = (0, 0)$. Evaluating Eq. \eqref{eq:agp_grad} is a straightforward calculation (see App. \ref{app:fixed_point} for details). We find

\begin{equation}
    \gradagpe = 
    \begin{bmatrix}
        \minus\frac{V'(0)}{\omega^2} & 0
    \end{bmatrix}
\end{equation}
The $q$-component of $\gradagpe$ vanishes, which happens in general when considering a time-reversible state and a perturbing function depending only on $q$. The $p$-component of $\gradagpe$ leads to $q_0\rightarrow 0 + \Delta\epsilon \frac{\partial\agpe}{\partial p} = - \Delta\epsilon\frac{V'(0)}{\omega^2}$, meaning the state simply shifts to the potential minimum of the perturbed Hamiltonian. On the other hand, for an unstable fixed point

\begin{equation}
    \ham = \frac{1}{2}\big(p^2 - \lambda^2 q^2\big) + \epsilon\ V(q)
    \label{eq:perturbed_upside_down_ho}
\end{equation}
we instead find that $\gradagpe\sim e^{\left(\frac{\lambda}{2\mu}\right)^2}$, and thus the fixed point can no longer be followed adiabatically.

\subsection{Linear Systems}

For Hamiltonians quadratic in the phase space variables $(p,q)$ with bounded motion, the AGP and its derivatives can be evaluated explicitly. Take for example a chain of harmonically coupled oscillators

\begin{equation}
    \ham =  \frac{1}{2}\sum_n\ p_n^2 + \Delta q_n^2
\end{equation}

\noindent where $\Delta q_n\equiv q_{n+1} - q_n$. The Hamiltonian can be rewritten in terms of complex mode coordinates

\begin{equation}
    \ham = \sum_k \omega_k\ a_k^* a_k
\end{equation}

\noindent where $a_k =\frac{1}{\sqrt{2 \omega_k}}\left(\omega_k Q_k + i P_k\right)$ and $(P_k, Q_k)$ are the linear normal modes. Now any perturbation can be written in the form
\begin{equation}
    \epsilon\ V(q) = \epsilon\ \sum_{\{k\}\{s\}} V_{k_1 k_2...} a^{s_1}_{k_1} a^{s_2}_{k_2}...
\end{equation}

\noindent where $s\in\{\pm1\}$ and the compact notation $a^{\minus1} =a^*$ is used. Then from Eq. \eqref{eq:agp_grad} we have

\begin{equation}
    \frac{\partial \agpe(a, a^*; \mu)}{\partial a^s_k} -i\ \sum_{\{k\}\{s\}} V_{k_1 k_2 ... k}\ \frac{a_{k_1}^{s_1}a_{k_2}^{s_2}...}{s_1 \omega_{k_1} + s_2 \omega_{k_2} + ... + s\ \omega_k}
\end{equation}

For example, take $V(q) = \frac{1}{3}\sum_n \Delta q_n^3$, which can be written 
\begin{equation}
    \frac{\alpha}{3} \sum_{\substack{k_1k_2k_3 \\ s_1 s_2 s_3} } V_{k_1 k_2 k_3} a_{k_1}^{s_1} a_{k_2}^{s_2} a_{k_3}^{s_3}
    \label{eq:alpha_fput_ca}
\end{equation}

\noindent then

\begin{equation}
    \frac{\partial \agpe(a, a^*; \mu)}{\partial a^s_k} = -i\ \sum_{\{k\}\{s\}} V_{k_1 k_2 k}\ \frac{a_{k_1}^{s_1}a_{k_2}^{s_2}}{s_1 \omega_{k_1} + s_2 \omega_{k_2} + s\ \omega_k}
\end{equation}

\noindent The corresponding deformation of the coordinates is 
\begin{equation}
\begin{aligned}
    &a_k \rightarrow a_k + \Delta\alpha\ \left( -i \frac{\partial\agpe}{\partial a_k^*}\right) \\ 
    &= a_k - \Delta\alpha \sum_{\{k\}\{s\}} V_{k_1 k_2 k}\ \frac{a_{k_1}^{s_1}a_{k_2}^{s_2}}{s_1 \omega_{k_1} + s_2 \omega_{k_2} - \omega_k} 
\end{aligned}
\end{equation}

\noindent the $\bigO(\Delta\alpha)$ deformation of $a_k$ is precisely the first-order correction given by canonical perturbation theory \cite{ShHe91}.

\section{Numerical Results}
In this section we present numerical results for systems covering a range of behaviors - the integrable Toda lattice, the Henon-Heiles system with mixed phase space, and the chaotic $\beta$-FPUT model. Numerics were performed using a fourth-order symplectic Runge-Kutta integrator \cite{CaSa1993} and time steps ranging from $\Delta t=0.01$ to $0.1$.

\label{sec:results}
\subsection{q-Breathers in the Toda and $\alpha$-FPUT lattices}

Here, we consider particular families of periodic orbits known as $q$-breathers\cite{FlIvKa06,KaRoCa24} in the Toda lattice \cite{To1967} described by Hamiltonian

\begin{equation}
    \ham_{\mathrm{Toda}} = \sum_n\ \frac{p_n^2}{2} + \frac{1}{4\alpha^2}e^{2\alpha\ \Delta q_n}
\end{equation}

\noindent as well as the $\alpha$-FPUT model \cite{FePaUlTs1955}

\begin{equation}
\begin{aligned}
    \ham_{\mathrm{FPUT}} &= \sum_n\  \frac{1}{2}\left(p_n^2 + \Delta q_n^2\right) + \frac{\alpha}{3} \Delta q_n^3\ \\ &\approx \ham_{\mathrm{Toda}} + \mathcal{O}(\alpha^2)
\end{aligned}
\end{equation}

Both models approach the linear chain as $\alpha\rightarrow 0$. We consider $q$-breathers with fixed energy and parametrized by the nonlinearity $\alpha$ such that for $\alpha=0$ the state is a linear normal mode. 

\begin{figure}[H] 
    \centering
    \includegraphics[width=0.6\textwidth]{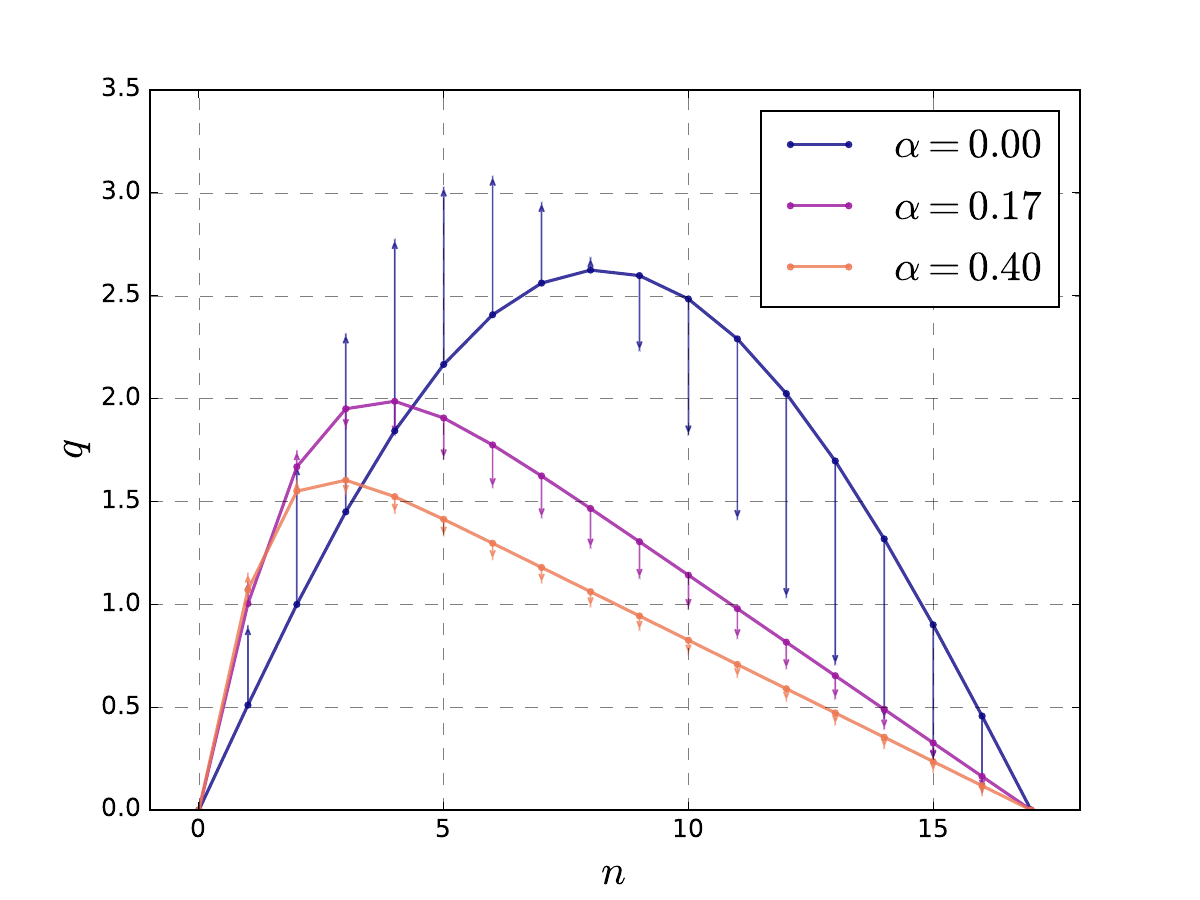} 
    \caption{Periodic orbits in the Toda lattice continued from a linear normal mode, for $N=16$ and nonlinearities $\alpha=0,\ 0.1, \text{and } 0.4$.  Oscillator positions are represented by dots and deformations generated by the AGP are shown as arrows, with an arbitrary but uniform scaling factor of $\Delta \alpha=0.05$.}
    \label{fig:toda_qbreather}
\end{figure}

In Fig. \ref{fig:toda_qbreather}, the $q$-breathers are plotted for the Toda lattice with $N=16$ sites, and energy $E=1.0$. The state initially in the longest wavelength normal mode deforms smoothly as a function of $\alpha$. The changes in coordinates generated by $\agp$, $q_n \rightarrow q_n +  \Delta \alpha\ \pb{\agp}{q_n}$ are shown as arrows, where an arbitrary scaling factor $\Delta \alpha = 0.05$ was used. Note that for each $\alpha$, the state is plotted at one of its two turning points so that all momenta $p_n$ vanish. As a result, only the $p$-directions of $\gradagp$ are nonzero.

The generated change of coordinates deforms the state more rapidly at small nonlinearities, where the initial sine wave grows steeper one one side and elongated on the other. The energy density of the state becomes more strongly localized in the left side of the lattice, where the difference in neighboring sites $\Delta q_n = q_{n+1} - q_n$ is large.

\begin{figure}[t!]
    \centering
    \hbox{\hspace{-35pt}
    \includegraphics[scale=.9]{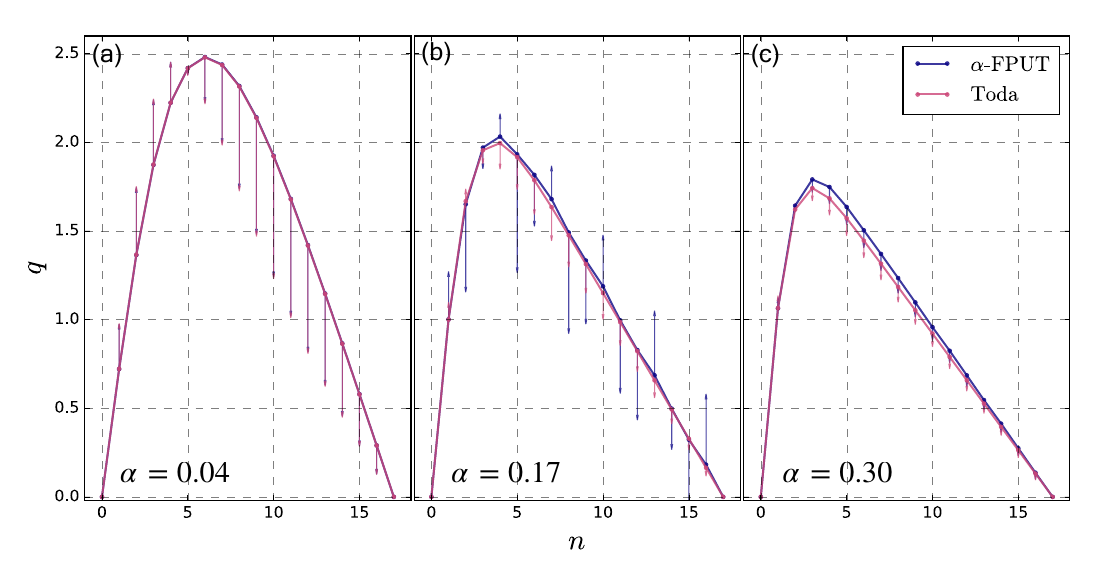}
   }
    \caption{Positions $q$ (dots) and AGP-generated deformations (arrows) as a function of site number $n$ for Toda and $\alpha$-FPUT systems, each with $N=16$ sites and energy $E=1.0$ and three different nonlinearities $\alpha=0.04,\ 0.17,$ and $0.30$.  The states of the $\alpha$-FPUT systems resemble those of the Toda lattice for each nonlinearity shown, but the deformations generated by the AGP differ strongly at the instability point $\alpha=0.17$ in panel (b).}
    \label{fig:alpha_toda_comp}
    \label{subfig:alpha_toda_comp_1}
    \label{subfig:alpha_toda_comp_2}
    \label{subfig:alpha_toda_comp_3}
    
\end{figure}

As a consequence of the Toda lattice's integrability, the transformation generated by the AGP is always finite. An interesting question is how robust is this result when integrability of the Hamiltonian is slightly broken. Now, we compare $\gradagp$ of the Toda lattice to the $\alpha$-FPUT model, whose Hamiltonian differs from Toda by terms of order $\bigO(\alpha^2)$ or smaller.

In Fig. \ref{fig:alpha_toda_comp} is shown the state and generated transformation for $q$-breathers in the Toda and $\alpha$-FPUT systems for $N=16$ sites and energy $E=1.0$. For small nonlinearities, the results for the two models nearly coincide as seen in Fig. \ref{subfig:alpha_toda_comp_1}. As the nonlinearity is varied, the $q$-breather in the $\alpha$-FPUT model is driven through an instability which may occur when two eigenvalues of Monodromy matrix collide \cite{KaRoCa24}. One such instability occurs around $\alpha=0.17$. The AGP gradient differs strongly at the instability point even though the states themselves are very similar, as can be seen in Fig. \ref{subfig:alpha_toda_comp_2}. Decreasing $\mu$ results in the magnitude $|\gradagp|$ of the transformation to grow exponentially. At higher nonlinearity the behavior between the two models becomes similar once again as shown in the third panel of Fig. \ref{subfig:alpha_toda_comp_3}.

\subsection{Henon-Heiles Model}
Next, we consider the Henon-Heiles model \cite{HeHe1964}, a well known system with mixed phase space, given by Hamiltonian

\begin{equation}
    \ham_{\mathrm{HH}} = \frac{1}{2}\left(p_x^2 + p_y^2 + x^2 + y^2\right) + \lambda\left(x^2y -\frac{y^3}{3}\right)
    \label{eq:henon_heiles}
\end{equation}
with $\lambda=1$. Above a dissociation energy $E=1/6$ the potential is no longer bounded from below, and trajectories become unbounded, whereas for small energies, the system is approximately harmonic and the phase space is full of regular trajectories. For intermediate values, a fraction of the phase space remains regular until the system becomes fully chaotic as $E$ approaches the dissociation energy \cite{Be1978}.

The Poincare section is a useful method to visualize dynamics in this low-dimensional system. We consider points on the plane $\mathcal{S}_y: \{(p_y, y)\ |\ x=0\}$. Since the phase space is four-dimensional and energy is conserved, a point $(p_y, y)\in\mathcal{S}_y$ fully determines the state of the system i.e. the $p_x$ component can be determined by $\ham(p_y, p_x, 0, y) =E$. 

\begin{figure}[t!]
    \centering
    \hbox{\hspace{-35pt}
    \includegraphics[scale=0.6]{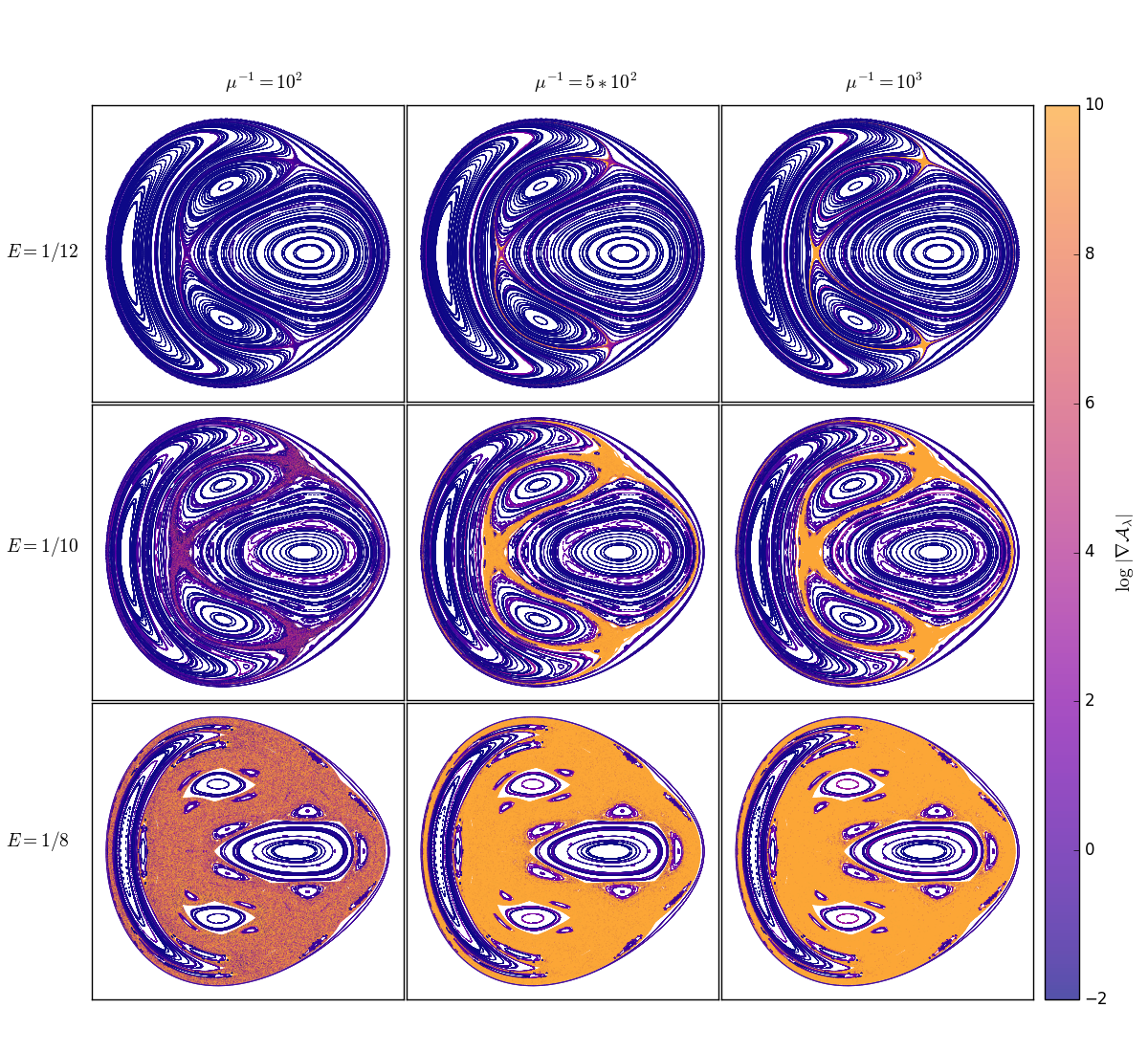}
    }
    \caption{Poincare sections of the Henon-Heiles model with varying energies $E=1/12,\ 1/10, $and $1/8$, and regularizer $\mu^{\minus 1} = 100,\ 500,\ $and $1000$. Vertical and horizontal axes represent $p_y$ and $y$ respectively, with ranges $p_y\in [- 0.48,\ 0.48]$ and $y\in [-0.42,\ 0.6]$.}
    \label{fig:HH_full_phase_space}
\end{figure}

In Fig. \ref{fig:HH_full_phase_space} the Poincare sections are plotted for the Henon-Heiles model at varying energies $E=1/12$, $1/10$, and $1/8$ as well as for different regularizers $\mu^{\minus 1} = 100,\ 500\ $ and $1000$. To create these plots, around $400$ initial conditions on the plane $\mathcal{S}_y$ at a given energy are chosen.  A mark is then placed at each point where the trajectory crosses the plane. Additionally, the gradient $\gradagp$ is computed at each of the initial points, and trajectories are color coded based on the log of the magnitude $|\gradagp(p_0, q_0)|$, with larger values being given brighter colors. The logarithm is taken to aid with visualizing the data.

Regions of phase space that have a nonzero Lyapunov exponent are evident from the magnitude of $\gradagp$ which spans several orders of magnitude. For $E=1/12$ the phase space is almost entirely regular, characterized by smooth, laminated curves formed from intersection points, and a small magnitude of $\gradagp$.

\begin{figure}[]
    \centering
    \includegraphics[width=0.6\textwidth]{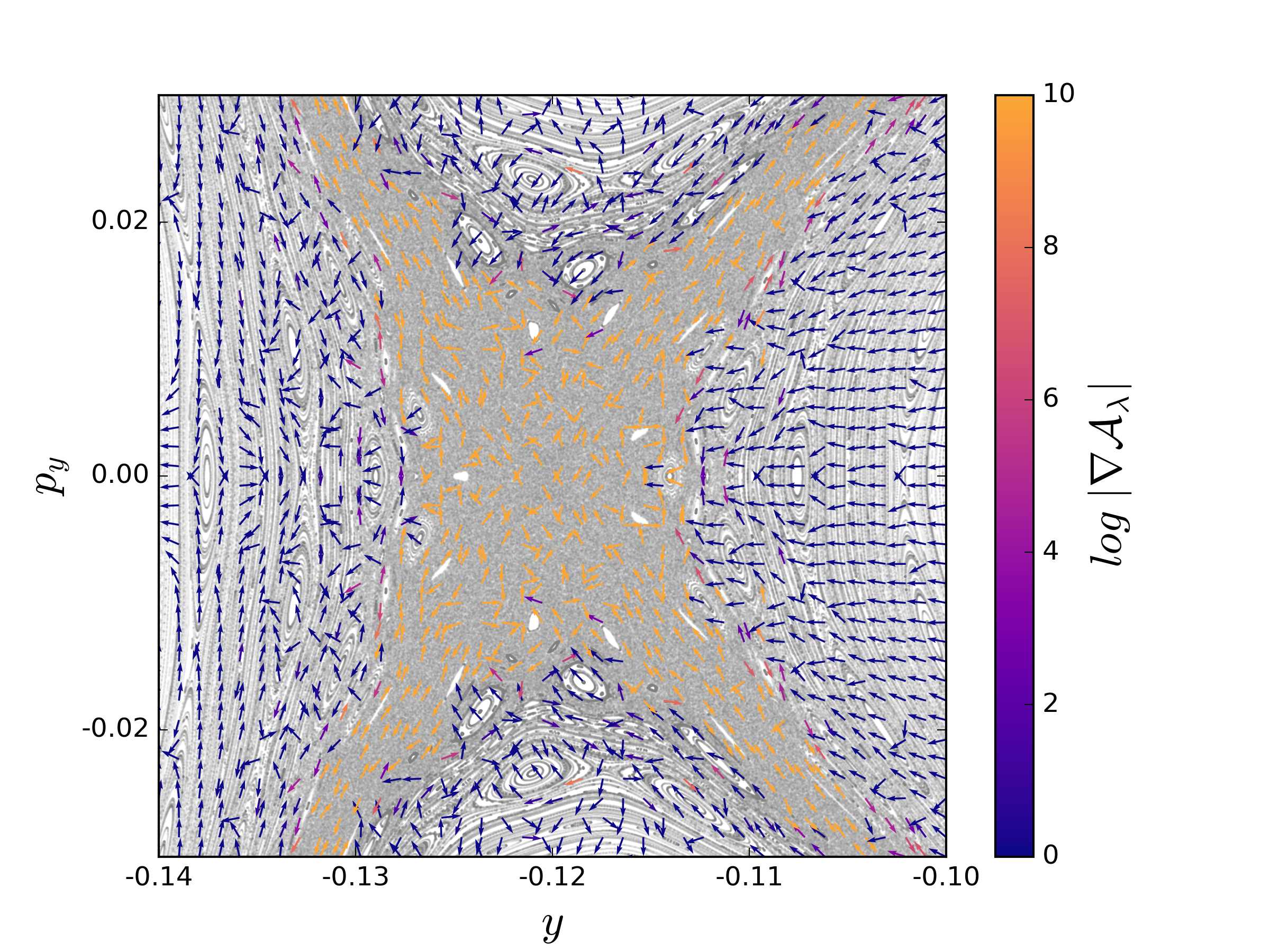} 
    \caption{Vector field $\Symp\gradagp$ with $\mu=10^{\minus 3}$ projected to the $(p_y, y)$ plane around a high sensitivity region in the Henon-Heiles model with low energy $E=1/12$. }
    \label{fig:hh_separatrix}
\end{figure}

In Fig. \ref{fig:hh_separatrix}, a closer look is taken at one of the unstable regions for $E=1/12$. Now the direction of the vector field corresponding to $\agp$ tangent to $\mathcal{S}_y$ is given by the direction of arrows, whose color again represents magnitude $|\gradagp|$. Obviously there is a massive separation in magnitude of the vector field between the regular and chaotic regions. Deep within the regular regions, the vector field is smooth, though it grows complex near the boundary of chaos.

\begin{figure}[]
    \centering
    \includegraphics[width=0.6\textwidth]{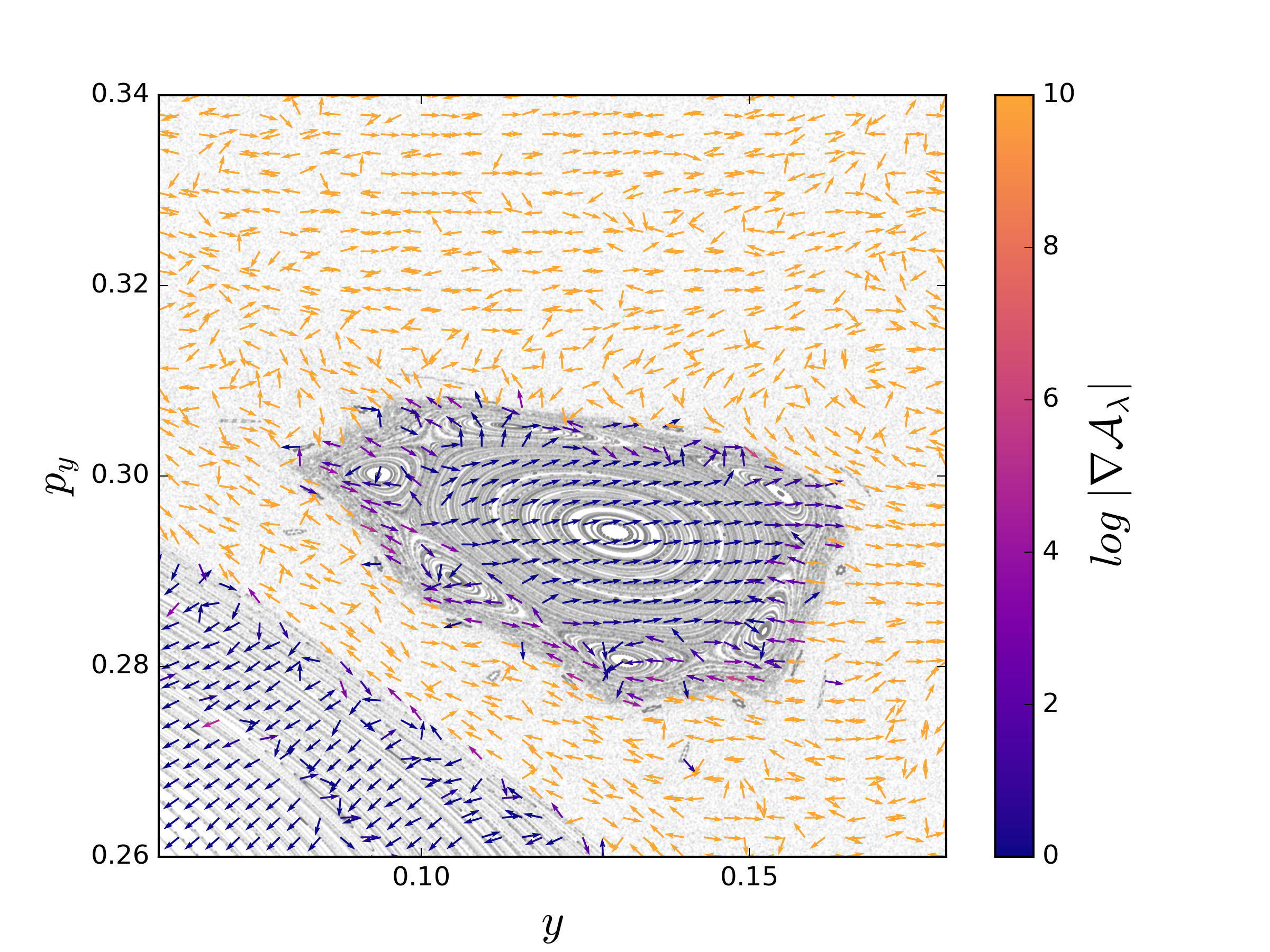} 
    \caption{Vector field $\Symp\gradagp$ with $\mu=10^{\minus 3}$ for the Henon-Heiles model with higher energy $E=1/8$. The zoomed in region is one of the regular islands in the bottom right panel of Fig. \ref{fig:HH_full_phase_space}. }
    \label{fig:hh_island}
\end{figure}

In Fig. \ref{fig:hh_island} the vector field is plotted around one of the regular islands appearing in Fig. \ref{fig:HH_full_phase_space} for $E=1/8$. At higher energy, the chaotic region has taken over most of the phase space, but some regular islands remain well insulated, giving pockets where the vector field $\Symp\gradagp$ is smooth and relatively uniform in space.

\subsection{$\beta$-FPUT Model}
Finally, we consider the $\beta$-FPUT \cite{FePaUlTs1955} model

\begin{equation}
    \ham = \frac{1}{2}\sum_n\left(p_n^2 + \Delta q_n^2\right) + \frac{\beta}{4}\Delta q_n^4
\end{equation}

which is nonintegrable and strongly chaotic for sufficiently large energy and nonlinearity $E\beta$.

\begin{figure}[H] 
    \centering
    \includegraphics[width=0.8\textwidth]{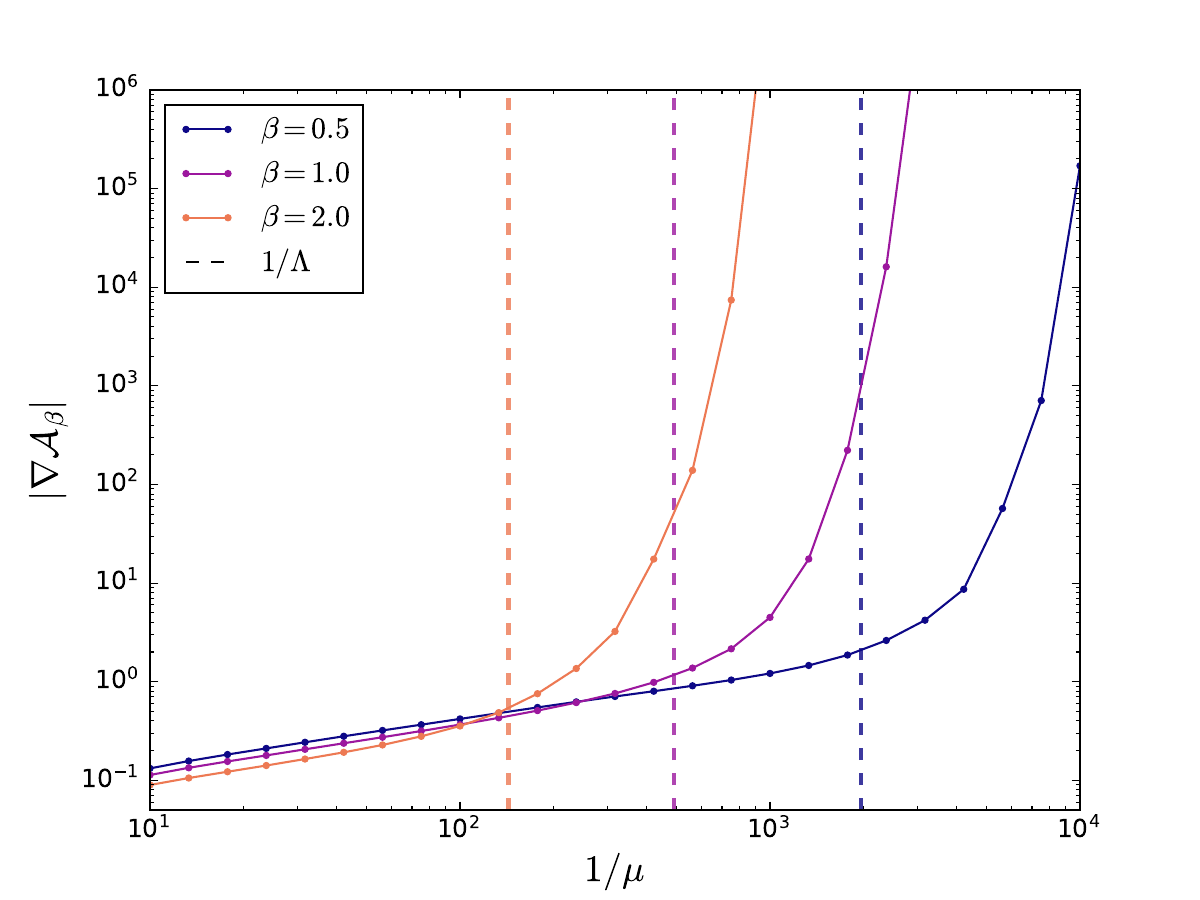} 
    \caption{Magnitude of the regularized AGP Gradient averaged over $100$ random initial states with energy $E=1$ in the $\beta$-FPUT model with $N=32$ sites. Dashed vertical lines show $1/\Lambda$ where $\Lambda$ is the maximal Lyapunov exponent, computed separately. The gradient of the AGP remains small approximately up until the Lyapunov time, after which it grows extremely rapidly.}
    \label{fig:grad_beta_fput}
\end{figure}

In Fig. \ref{fig:grad_beta_fput} is shown the magnitude of $\gradagpb$ averaged over $100$ initial conditions obtained by generating a random state with energy $E=1.0$ spread uniformly over the $N=32$ normal modes. The gradient is computed over a range of regularizers $\mu$. Three values of the nonlinearity $\beta=0.5,\ 1.0,$ and $2.0$ are considered, and the MLEs $\Lambda$ are computed separately using a standard method \cite{BeGaGiSt1980}.

The computed magnitudes $|\gradagpb(\mu)|$ diverge for earlier $\mu^{\minus1}$ with increasing nonlinearity, and at times approximately $\mu\sim\Lambda$. For small nonlinearity, it is known that the $\beta$-FPUT model exhibits extremely slow thermalization. Understanding how these two timescales are related remains an interesting open problem.

Next, we probe how specific directions of $\gradagpb$ corresponding to CLVs depend on $\mu$. To compute this we compute Eq. \eqref{eq:agp_grad} but with columns of the initial matrix $\Phi(0)$ initialized as unit Lyapunov vectors $v_j$. The dynamics conserve these subspaces, meaning that a vector initialized as a CLV $v_j$ at time $t=0$ will evolve to remain parallel to the CLV at a later time $v_j(T)$. The growth rate is given by the corresponding finite time Lyapunov exponent $\lambda_j(p,q;T)$, which approaches the true Lyapunov exponent $\lambda_j$ as $T\rightarrow\infty$. From Eq. \eqref{eq:agp_grad}, directions $\partial\agpb(\mu)$ should therefore diverge exponentially for $\mu<\lambda_j$, implying that for intermediate values of $\mu$ there may be a very large variation among directions of $\gradagpb$.

\begin{figure}[H] 
    \centering
    \includegraphics[width=0.8\textwidth]{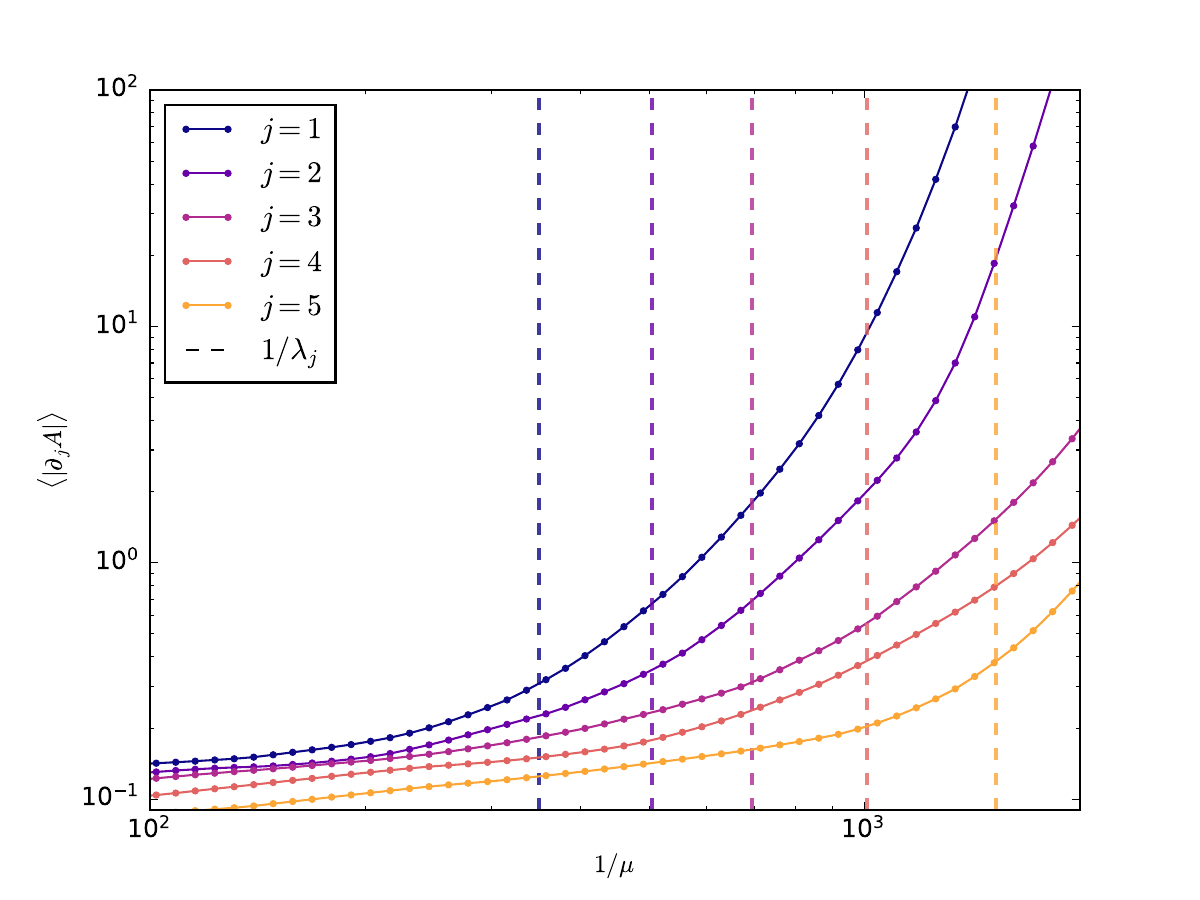} 
    \caption{Absolute values of particular components of $\gradagpb$ taken in directions of the first few covariant Lyapunov vectors $v_j$ and averaged over $100$ points on a single trajectory, for the $\beta$-FPUT model with $N=8$.}
    \label{fig:beta_clv_grad}
\end{figure}

Variations of the local finite-time Lyapunov exponents $\lambda_j(p,q; T)$ cause the separation between directions at individual points to be somewhat noisy, and so we average the magnitude $|\partial_j\agpb(p,q;\mu)|$ over $100$ points $(p,q)$ sampled along a trajectory. Absolute values of components $\partial_j\agpb(p,q;\mu)$ are taken prior to averaging, as in general the signs of these terms can be positive or negative.

We obtain averaged $\gradagpb(p, q; \mu)$ in the directions of the first few CLVs $v_j$ for $j=1-5$ for the $\beta$-FPUT model with $N=8$ sites and $\beta=0.5$, shown in Fig. \ref{fig:beta_clv_grad}. We find that $\gradagpb$ in directions of $v_j$ diverge at progressively larger times $\mu^{\minus 1}$. Dashed vertical lines show the corresponding inverse Lyapunov exponents $\lambda_j^{\minus1}$.

This approach as straightforwardly implemented is limited by exponential instabilities, making it difficult to continue to smaller $\mu$ which involves integrating for a longer time. Nevertheless, a large hierarchy in magnitudes between various directions of $\gradagpb$ corresponding to the CLVs is observed as predicted.

\section{Conclusions}
\label{sec:discussion}
In this paper, we have studied the AGP and its gradient in classical systems. We formulate Eq. \ref{eq:agp_grad} for the computation of the regularized AGP gradient dependent upon a time cutoff $\mu$. The regularized AGP yields phase space functions which generate canonical transformations via taking the Poisson bracket with the coordinate functions. As the Poisson bracket involves derivatives of a function with respect to the complete set of phase space coordinates, the gradient of the AGP as a classical function is crucial to understanding these transformations.

The method introduced in this paper involves a direct calculation of a trajectory and the linearized flow around it. A consequence of the sensitive dependence of initial conditions in chaotic systems is that the gradient of the AGP becomes unbounded when $\mu$ is smaller than the MLE. Thus, we conclude that the adiabatic transformation generated by the AGP remains valid only up to the Lyapunov time. This result is supported by numerical examples, and contrasted with integrable examples where the gradient remains bounded.

 Our method involves an exact computation, in contrast with various approximate methods which have been developed. Approximations of the AGP using only local terms have the greatest promise for experimental applications. Thus it would be interesting to compare exact results for the AGP and its gradient to results of approximate methods. We leave this study for future work.

As discussed earlier, the gradient of the AGP diverges when there is a positive Lyapunov exponent i.e in chaotic systems. Nevertheless, the Poisson bracket of the AGP and a phase space distribution $\rho$ may still finite, which places constraints on the form of $\rho$.  Indeed, stationary distributions in chaotic systems cover an energy shell uniformly, which guarantees that the Poisson bracket with the regularized AGP is finite. The sensitivity of a distribution to parametric change can be quantified by Fisher information, which has already been used to describe chaos in quantum systems and is closely related to the AGP norm. Exploring this connection further is another promising direction of future work. 

\section*{Acknowledgements}
The authors would like to thank Anatoli Polkovnikov and Dries Sels for helpful discussions. We would also like to acknowledge use of Boston University's Shared Computing Cluster (SCC).

\bibliography{bib.bib}

\begin{thebibliography}{10}
\providecommand{\url}[1]{\texttt{#1}}
\providecommand{\urlprefix}{URL }
\expandafter\ifx\csname urlstyle\endcsname\relax
  \providecommand{\doi}[1]{doi:\discretionary{}{}{}#1}\else
  \providecommand{\doi}{doi:\discretionary{}{}{}\begingroup \urlstyle{rm}\Url}\fi
\providecommand{\eprint}[2][]{\url{#2}}

\bibitem{Ka1950}
T.~Kato,
\newblock \emph{On the adiabatic theorem of quantum mechanics},
\newblock Journal of the Physical Society of Japan \textbf{5}(6), 435 (1950),
\newblock \doi{10.1143/JPSJ.5.435},
\newblock \eprint{https://doi.org/10.1143/JPSJ.5.435}.

\bibitem{KoSeMePo17}
M.~Kolodrubetz, D.~Sels, P.~Mehta and A.~Polkovnikov,
\newblock \emph{Geometry and non-adiabatic response in quantum and classical systems},
\newblock Physics Reports \textbf{697}, 1 (2017),
\newblock \doi{https://doi.org/10.1016/j.physrep.2017.07.001},
\newblock Geometry and non-adiabatic response in quantum and classical systems.

\bibitem{DeRi03}
M.~Demirplak and S.~A. Rice,
\newblock \emph{Adiabatic population transfer with control fields},
\newblock The Journal of Physical Chemistry A \textbf{107}(46), 9937 (2003),
\newblock \doi{10.1021/jp030708a},
\newblock \eprint{https://doi.org/10.1021/jp030708a}.

\bibitem{Be09}
M.~V. Berry,
\newblock \emph{Transitionless quantum driving},
\newblock Journal of Physics A: Mathematical and Theoretical \textbf{42}(36), 365303 (2009),
\newblock \doi{10.1088/1751-8113/42/36/365303}.

\bibitem{Ja2013}
C.~Jarzynski,
\newblock \emph{Generating shortcuts to adiabaticity in quantum and classical dynamics},
\newblock Phys. Rev. A \textbf{88}, 040101 (2013),
\newblock \doi{10.1103/PhysRevA.88.040101}.

\bibitem{dCa13}
A.~del Campo,
\newblock \emph{Shortcuts to adiabaticity by counterdiabatic driving},
\newblock Phys. Rev. Lett. \textbf{111}, 100502 (2013),
\newblock \doi{10.1103/PhysRevLett.111.100502}.

\bibitem{HoRaJa25}
R.~Holtzman, O.~Raz and C.~Jarzynski,
\newblock \emph{Shortcuts to adiabaticity across a separatrix},
\newblock Phys. Rev. Lett. \textbf{134}, 157201 (2025),
\newblock \doi{10.1103/PhysRevLett.134.157201}.

\bibitem{PaClCaPoSe20}
M.~Pandey, P.~W. Claeys, D.~K. Campbell, A.~Polkovnikov and D.~Sels,
\newblock \emph{Adiabatic eigenstate deformations as a sensitive probe for quantum chaos},
\newblock Phys. Rev. X \textbf{10}, 041017 (2020),
\newblock \doi{10.1103/PhysRevX.10.041017}.

\bibitem{SePo21}
D.~Sels and A.~Polkovnikov,
\newblock \emph{Dynamical obstruction to localization in a disordered spin chain},
\newblock Phys. Rev. E \textbf{104}, 054105 (2021),
\newblock \doi{10.1103/PhysRevE.104.054105}.

\bibitem{SePo23}
D.~Sels and A.~Polkovnikov,
\newblock \emph{Thermalization of dilute impurities in one-dimensional spin chains},
\newblock Phys. Rev. X \textbf{13}, 011041 (2023),
\newblock \doi{10.1103/PhysRevX.13.011041}.

\bibitem{LiMaKiPoFl24}
C.~Lim, K.~Matirko, H.~Kim, A.~Polkovnikov and M.~O. Flynn,
\newblock \emph{Defining classical and quantum chaos through adiabatic transformations} (2024), \eprint{2401.01927}.

\bibitem{SwLyVi25}
R.~{{\'S}wi{\k{e}}tek}, P.~{{\L}yd{\.z}ba} and L.~{Vidmar},
\newblock \emph{{Fading ergodicity meets maximal chaos}},
\newblock arXiv e-prints arXiv:2502.09711 (2025),
\newblock \doi{10.48550/arXiv.2502.09711},
\newblock \eprint{2502.09711}.

\bibitem{KaRoCa25}
N.~Karve, N.~Rose and D.~Campbell,
\newblock \emph{Adiabatic gauge potential as a tool for detecting chaos in classical systems} (2025), \eprint{2502.12046}.

\bibitem{BaArChPo25}
B.~Barrera, D.~P. Arovas, A.~Chandran and A.~Polkovnikov,
\newblock \emph{The moving born-oppenheimer approximation} (2025), \eprint{2502.17557}.

\bibitem{SePo17}
D.~Sels and A.~Polkovnikov,
\newblock \emph{Minimizing irreversible losses in quantum systems by local counterdiabatic driving},
\newblock Proceedings of the National Academy of Sciences \textbf{114}(20), E3909 (2017),
\newblock \doi{10.1073/pnas.1619826114}.

\bibitem{GjCaPo22}
N.~O. Gjonbalaj, D.~K. Campbell and A.~Polkovnikov,
\newblock \emph{Counterdiabatic driving in the classical $\ensuremath{\beta}$-fermi-pasta-ulam-tsingou chain},
\newblock Phys. Rev. E \textbf{106}, 014131 (2022),
\newblock \doi{10.1103/PhysRevE.106.014131}.

\bibitem{TadCa24}
K.~Takahashi and A.~del Campo,
\newblock \emph{Shortcuts to adiabaticity in krylov space},
\newblock Phys. Rev. X \textbf{14}, 011032 (2024),
\newblock \doi{10.1103/PhysRevX.14.011032}.

\bibitem{Bh23}
B.~Bhattacharjee,
\newblock \emph{A lanczos approach to the adiabatic gauge potential} (2023), \eprint{2302.07228}.

\bibitem{FiNoCaLuSe25}
J.~R. Finžgar, S.~Notarnicola, M.~Cain, M.~D. Lukin and D.~Sels,
\newblock \emph{Counterdiabatic driving with performance guarantees} (2025), \eprint{2503.01958}.

\bibitem{MoPo25}
S.~Morawetz and A.~Polkovnikov,
\newblock \emph{Universal counterdiabatic driving} (2025), \eprint{2503.01952}.

\bibitem{KaRoCa2025_2}
N.~Karve, N.~Rose and D.~Campbell,
\newblock \emph{Diffusion as a signature of chaos} (2025), \eprint{2507.18617}.

\bibitem{KuGrPr17}
I.~{Kukuljan}, S.~{Grozdanov} and T.~{Prosen},
\newblock \emph{{Weak quantum chaos}},
\newblock Phys. Rev. B \textbf{96}(6), 060301 (2017),
\newblock \doi{10.1103/PhysRevB.96.060301},
\newblock \eprint{1701.09147}.

\bibitem{GuKiZh22}
Y.~{Gu}, A.~{Kitaev} and P.~{Zhang},
\newblock \emph{{A two-way approach to out-of-time-order correlators}},
\newblock Journal of High Energy Physics \textbf{2022}(3), 133 (2022),
\newblock \doi{10.1007/JHEP03(2022)133},
\newblock \eprint{2111.12007}.

\bibitem{Tr23}
D.~A. Trunin,
\newblock \emph{Refined quantum lyapunov exponents from replica out-of-time-order correlators},
\newblock Phys. Rev. D \textbf{108}, 105023 (2023),
\newblock \doi{10.1103/PhysRevD.108.105023}.

\bibitem{PaCaAvScAl19}
D.~E. Parker, X.~Cao, A.~Avdoshkin, T.~Scaffidi and E.~Altman,
\newblock \emph{A universal operator growth hypothesis},
\newblock Phys. Rev. X \textbf{9}, 041017 (2019),
\newblock \doi{10.1103/PhysRevX.9.041017}.

\bibitem{RoBe1992}
J.~M. Robbins and M.~V. Berry,
\newblock \emph{The geometric phase for chaotic systems},
\newblock Proceedings of the Royal Society of London. Series A: Mathematical and Physical Sciences \textbf{436}, 631 (1992).

\bibitem{Ja1995}
C.~Jarzynski,
\newblock \emph{Geometric phases and anholonomy for a class of chaotic classical systems},
\newblock Phys. Rev. Lett. \textbf{74}, 1732 (1995),
\newblock \doi{10.1103/PhysRevLett.74.1732}.

\bibitem{CaSa1993}
M.~P. Calvo and J.~M. Sanz-Serna,
\newblock \emph{The development of variable-step symplectic integrators, with application to the two-body problem},
\newblock SIAM Journal on Scientific Computing \textbf{14}(4), 936 (1993),
\newblock \doi{10.1137/0914057},
\newblock \eprint{https://doi.org/10.1137/0914057}.

\bibitem{YaSt75}
V.~A. Yakubovich and V.~M. Starzhinskii,
\newblock \emph{Linear Differential Equations with Periodic Coefficients}, vol.~1,
\newblock John Wiley \& Sons,
\newblock Translated from Russian by D. Louvish (1975).

\bibitem{FlIvKa06}
S.~Flach, M.~V. Ivanchenko and O.~I. Kanakov,
\newblock \emph{$q$-breathers in fermi-pasta-ulam chains: Existence, localization, and stability},
\newblock Phys. Rev. E \textbf{73}, 036618 (2006),
\newblock \doi{10.1103/PhysRevE.73.036618}.

\bibitem{GiPoTuChPo2007}
F.~Ginelli, P.~Poggi, A.~Turchi, H.~Chat\'e, R.~Livi and A.~Politi,
\newblock \emph{Characterizing dynamics with covariant lyapunov vectors},
\newblock Phys. Rev. Lett. \textbf{99}, 130601 (2007),
\newblock \doi{10.1103/PhysRevLett.99.130601}.

\bibitem{KuPa2012}
P.~V. Kuptsov and U.~Parlitz,
\newblock \emph{Theory and computation of covariant lyapunov vectors},
\newblock Journal of Nonlinear Science \textbf{22}(5), 727 (2012),
\newblock \doi{10.1007/s00332-012-9126-5}.

\bibitem{ha10}
M.~B. Hastings,
\newblock \emph{Quasi-adiabatic continuation for disordered systems: Applications to correlations, lieb-schultz-mattis, and hall conductance} (2010), \eprint{1001.5280}.

\bibitem{ShHe91}
D.~S. Sholl and B.~I. Henry,
\newblock \emph{Recurrence times in cubic and quartic fermi-pasta-ulam chains: A shifted-frequency perturbation treatment},
\newblock Phys. Rev. A \textbf{44}, 6364 (1991),
\newblock \doi{10.1103/PhysRevA.44.6364}.

\bibitem{KaRoCa24}
N.~Karve, N.~Rose and D.~Campbell,
\newblock \emph{Periodic orbits in fermi–pasta–ulam–tsingou systems},
\newblock Chaos: An Interdisciplinary Journal of Nonlinear Science \textbf{34}(9), 093117 (2024),
\newblock \doi{10.1063/5.0223767},
\newblock \eprint{https://pubs.aip.org/aip/cha/article-pdf/doi/10.1063/5.0223767/20161085/093117\_1\_5.0223767.pdf}.

\bibitem{To1967}
M.~{Toda},
\newblock \emph{{Vibration of a Chain with Nonlinear Interaction}},
\newblock Journal of the Physical Society of Japan \textbf{22}(2), 431 (1967),
\newblock \doi{10.1143/JPSJ.22.431}.

\bibitem{FePaUlTs1955}
E.~Fermi, J.~Pasta, S.~Ulam and M.~Tsingou,
\newblock \emph{Studies of the nonlinear problems},
\newblock Tech. rep.,
\newblock \doi{10.2172/4376203} (1955).

\bibitem{HeHe1964}
M.~{Henon} and C.~{Heiles},
\newblock \emph{{The applicability of the third integral of motion: Some numerical experiments}},
\newblock Astronomical Journal \textbf{69}, 73 (1964),
\newblock \doi{10.1086/109234}.

\bibitem{Be1978}
M.~V. Berry,
\newblock \emph{Regular and irregular motion},
\newblock In S.~Jorna, ed., \emph{Topics in Nonlinear Mechanics}, vol.~46 of \emph{American Institute of Physics Conference Proceedings}, pp. 16--120 (1978).

\bibitem{BeGaGiSt1980}
G.~Benettin, L.~Galgani, A.~Giorgilli and J.-M. Strelcyn,
\newblock \emph{Lyapunov characteristic exponents for smooth dynamical systems and for hamiltonian systems; a method for computing all of them. part 2: Numerical application},
\newblock Meccanica \textbf{15}, 21 (1980),
\newblock \doi{10.1007/BF02128237}.

\bibitem{Os1968}
V.~I. Oseledets,
\newblock \emph{A multiplicative ergodic theorem. ljapunov characteristic numbers for dynamical systems},
\newblock Trudy Moskov. Mat. Obsc. \textbf{19}, 179 (1968),
\newblock English transl. in Trans. Moscow Math. Soc., 19 (1968), 197--231.

\bibitem{Ru1979}
D.~Ruelle,
\newblock \emph{Ergodic theory of differentiable dynamical systems},
\newblock Publications Math\'ematiques de l'IH\'ES \textbf{50}, 27 (1979).

\bibitem{GiChLiPo13}
F.~Ginelli, H.~Chaté, R.~Livi and A.~Politi,
\newblock \emph{Covariant lyapunov vectors} \textbf{46}(25), 254005 (2013),
\newblock \doi{10.1088/1751-8113/46/25/254005}.

\end{thebibliography}

\section{Appendices}
\begin{appendices}
\numberwithin{equation}{section}

\section{Gaussian Regularization} 
\label{app:regularization}
Equation \eqref{eq:agp_grad} diverges for $\mu < \Lambda$, where $\Lambda$ is the MLE of the system. This implies that in chaotic systems the AGP becomes discontinuous after a finite time. When evaluating Eq. \eqref{eq:agp_grad} numerically however, integration can only be performed over a finite time $T$, resulting in a very large but finite value that depends noisily on the cutoff time and may be subject to numerical overflow. Therefore, one can replace Eq. \eqref{eq:agp_grad} with
\begin{equation}
    \gradagp(p,q; \mu) = -\frac{1}{2} \int_{\minus \infty}^{\infty}dt\ \text{sgn}(t)\ \exp({\minus (\mu t)^2})\ \nabla V(t)\cdot\Phi(t) 
    \label{eq:agp_grad_gauss}
\end{equation}

which still grows large but in a more controlled way, and which theoretically is always finite for finite $\mu$.

To analyze the behavior of Eq. \eqref{eq:agp_grad_gauss}, consider the following form of the integrand $\nabla V(t)\cdot\Phi(t) \sim e^{\Lambda t + i\omega t}$

\begin{equation}
    \gradagp(p,q; \mu) \sim -\frac{1}{2} \int_{\minus \infty}^{\infty} dt\ \text{sgn}(t)\ \exp(\minus \mu^2 t^2 + (\Lambda + i \omega) t)
\end{equation}

completing the square in the exponent and then evaluating the Gaussian integrals gives

\begin{equation}
    \frac{\sqrt{\pi}}{\mu} erf\bigg(\frac{z}{2\mu}\bigg)e^{\big(\frac{z}{2\mu}\big)^2}
    \label{eq:erf}
\end{equation}
with $z=\Lambda + i\omega$. This expression scales as $e^{\big(\frac{\Lambda}{2\mu}\big)^2}$ when $\Lambda$ is nonzero. On the other hand, if $\Lambda=0$ the $z = i\omega$ is pure imaginary. In this case

\begin{equation}
    \lim_{\mu\rightarrow0} erf\bigg(\frac{i\omega}{2\mu}\bigg) = \frac{\mu}{\sqrt{\pi}}\ \frac{e^{\big(\frac{\omega}{2\mu}\big)^2}}{\omega}
\end{equation}
and the expression Eq. \eqref{eq:erf} becomes $1/\omega$ in agreement with the standard regularization procedure for $\mu\rightarrow0$.
\section{Fixed Point Calculation}
\label{app:fixed_point}
The tangent evolution is solved by diagonalizing the matrix $\Phi(t) = PD(t)P^{-1}$ where $D(t) = diag(e^{zt}, e^{\minus zt})$ with $z, \minus z$ the eigenvalues of $J$, and the columns of $P$ are the eigenvectors of $J$.  For the stable fixed point $z=i\omega$. Eq. \eqref{eq:agp_grad_gauss} gives

\begin{equation}
    \gradagpe = -\frac{1}{2}\ \nabla V \cdot P\ \bigg[ \int_0^{\infty} dt\ e^{\minus \mu^2t^2}\ \text{diag}(e^{\pm i \omega t})  - \int_{\minus\infty}^{0} dt\ e^{\minus \mu^2t^2}\ \text{diag}(e^{\pm i \omega t})\bigg]\ P^{\minus 1}
    \label{eq:ex_grad_agp_fp}
\end{equation}
The integrals can be evaluated and their difference gives $\frac{\sqrt{\pi}}{\mu}erf\left(\frac{\pm i\omega}{2\mu}\right)e^{\minus \left(\frac{\omega}{2\mu}\right)^2}$, which in the limit $\mu\rightarrow0$ becomes $\pm i/\omega$. Using
\begin{equation}
    P = \frac{1}{\sqrt{2}}
\begin{pmatrix}
1 & 1 \\
i/\omega & \minus i/\omega
\end{pmatrix},\ P^{\minus 1} =
\frac{1}{\sqrt{2}}
\begin{pmatrix}
1 & \minus i\omega \\
1 & i\omega
\end{pmatrix}
\end{equation}
and evaluating Eq. \eqref{eq:ex_grad_agp_fp} we find

\begin{equation}
    \gradagpe = 
    \begin{bmatrix}
        \minus\frac{V'(0)}{\omega^2} & 0
    \end{bmatrix}
\end{equation}

\section{Covariant Lyapunov Vectors}
\label{app:clvs}
Covariant Lyapunov vectors define subspaces of the tangent space that are invariant in time \cite{Os1968, Ru1979, GiPoTuChPo2007, KuPa2012}.  If the Lyapunov spectrum $\lambda_1 > \lambda_2 >\ ...\ >\lambda_N=0$ is nondegenerate i.e. each $\lambda_j$ is distinct, then these subspaces are one-dimensional and can each be specified by a unit vector $v_j$. If one or more $\lambda_j$ are equal, then the corresponding subspace will be of greater dimension. A tangent vector $\xi$ initialized along some $v_j(p_0, q_0)$ will evolve over time $t$ to $e^{\lambda_j(p_0, q_0;\ t)t}$, where the finite-time Lyapunov exponent $\lambda(p_0, q_0; t)$ describes the growth rate of tangent vectors.

The method of computing CLVs given by Ginelli et al. \cite{GiChLiPo13} is an extension of the widely used algorithm of by Benettin et al.\cite{BeGaGiSt1980} for computing the Lyapunov spectrum. The Lyapunov spectrum is defined as the eigenvalues of the matrix

\begin{equation}
    \lim_{\tau\rightarrow\infty} \frac{1}{2\tau} \log \Phi(\tau)^T\Phi(\tau)
\end{equation}
where $\Phi(0) = \Id$. Since a generic tangent vector grows exponentially with time, the values of $\Phi(t)$ grow very large, and more importantly the columns become nearly parallel, leading to a very poorly conditioned matrix for numerical treatment.

The algorithm of Benettin et al. handles this by re-orthonormalizing the matrix $\Phi(t)$ periodically via a Gram-Schmidt process, yielding $\Phi(t) = QR$. The matrix $\Phi$ is then set equal to the orthogonal $Q$ and the diagonal entries of the upper triangular matrix $R$ give the local growth rates. Under alternated evolution and orthogonalization, the generated matrices $Q$ converge to a sequence known as the Gram-Schmidt (GS) basis.

Ginelli et al.'s algorithm for computing the CLVs consists of iterating backwards in time a matrix of coefficients $C_k$ expressing the CLVs in the GS basis

\begin{equation}
    C_{k} = R_k^{\minus 1} C_{k+1} D_k
    \label{eq:ginelli_it}
\end{equation}
Where $\{R_k\}$ is the sequence of matrices generated from from the Benettin et al. algorithm and diagonal matrix $D_k \equiv Diag(R_k)$ contains the growth rates. Equation \eqref{eq:ginelli_it} effectively undoes the changes effected on the evolved tangent vectors by the orthogonalization process, while keeping the coefficient matrix $C_k$ of roughly constant order through multiplication by $D_k$. Finding the CLVs in the coordinate basis is then done by expanding $C_k$ in the GS basis: $V_k = Q_kC_k$.

Ginelli et al.'s algorithm requires evolving the state plus tangent vectors through a forward transient time $\tau_1$ for the orthogonal matrices $Q$ to converge to the GS basis, and iterating Eq. \eqref{eq:ginelli_it} through a backward transient time $\tau_2$. To compute the CLV basis for Fig. \ref{fig:beta_clv_grad} equal times $\tau_1=\tau_2=10^6$ and an integration time step of $dt=0.01$ were used. As a check, we verified that the computed Lyapunov spectrum satisfied $|\lambda_{\minus j}| =\lambda_{j}$ to well within $1\%$ for each $j$. The smallest pair of values $\lambda_{\pm N}$ were not exactly zero due to finite integration time, but were around $4*10^{-6}$.

\end{appendices}

\end{document}